# Application of RESNET50 Convolutional Neural Network for the Extraction of Optical Parameters in Scattering Media.

B Deng*[1], Y Zhang*[2] , A Parkes[1], A Bentley[2], A J Wright[2], M Pound[1] and M G Somekh[2]

- These authors contributed equally to this work

[1] School of Computer Science, University of Nottingham, NG8 1BB, UK

[2] Optics and Photonics Group, Faculty of Engineering, University of Nottingham, Nottingham, NG7 2RD, UK

**Abstract:**

Estimation of the optical properties of scattering media such as tissue is important in diagnostics as well as in the development of techniques to image deeper. As light penetrates the sample scattering events occur that alter the propagation direction of the photons in a random manner leading degradation of image quality. The distribution of the scattered light does, however, give a measure of the optical properties such as the reduced scattering coefficient and the absorption coefficient. Unfortunately, inverting scattering patterns to recover the optical properties is not simple especially in the regime where the light is partially randomized. Machine learning has been proposed by several authors as a means of recovering these properties from either the back scattered or the transmitted light. In the present paper we train a general purpose convolutional neural network RESNET 50 with simulated data based on Monte Carlo simulations. We show that compared with previous work our approach gives comparable or better reconstruction accuracy with training on a much smaller dataset. Moreover, by training on multiple parameters such as the intensity distribution at multiple planes or the exit angle and spatial distribution one achieves improved performance compared to training on a single input such as the intensity distribution captured at the sample surface. While our approach gives good parameter reconstruction, we identify factors that limit accuracy of the recovered properties, particularly the absorption coefficient. In the light of these limitations, we suggest how the present approach may be enhanced for even better performance.

**Introduction:**

The optical properties of biological tissues play a vital role in medical and biological diagnostic and therapeutic applications. When tissue interacts with light it is scattered or randomized as it propagates and, furthermore, when absorption of the light takes place the incoming photons are annihilated. Knowledge of the detailed properties of light scattering is very important if one wants to 'unscramble' the light; that is to restore, for instance, a sharp focus within the tissue. The effect of absorption is, naturally, not reversible, but, nevertheless, knowledge of its value is important to understand limitations on, for instance, penetration depth. While one may want to correct for the effect of tissue scattering there are cases where one simply wants to be able to characterize the tissue properties so that, one can make a distinction between different tissue types and pathologies [1]. From the point of view of fundamental optical physics measuring the scattering properties is important such as in the determination of the range of the angle memory effect [2] and the regions of validity of the diffusion approximation [3]. Unfortunately, a recent review by our group looking at the optical properties of skin has shown that the literature is inconsistent between different authors and in many cases the conditions under which the data was obtained is often ill-defined [4].

Tissue can be characterized by three key optical properties the scattering coefficient, $\mu_s$ , absorption coefficient, $\mu_a$, and anisotropy factor, $g$. As a light beam propagates through tissue, the scattering coefficient describes the number of scattering events that occur per unit distance, a scattering event is when the propagation direction of a photon is altered due to the presence of a scattering center. Different scattering events have different effects and this may be represented by the anisotropy coefficient which is defined as the mean of the cosine of the angular deviation caused by the scattering. A value of $g = 0$ implies isotropic scattering, whereas values close to 1 indicates that the scattering is mainly in the forward direction, for tissue this value is typically between 0.8 and 0.99 [4]. Numbers close to 1 mean that on average each scattering event introduces only a small deviation in the direction of the photons. When many random scattering events occur the direction of the light becomes randomized, when this occurs the values of $\mu_s$ and $g$ cannot be separated and are described with a single parameter, the reduced scattering coefficient, $\mu_s' = \mu_s(1 - g)$. In essence, this is saying that when the details of the scattering events are lost due to randomization the effect of scattering can be bundled into a single parameter which is the product of the number of scattering events and their strength, $1 - g$. The absorption coefficient gives the absorption per unit distance and the number of photons decays as $\exp -\mu_a x$ where $x$ is the propagation distance.

There are several methods to measure the optical properties of a tissue. For *ex-vivo* measurement, the Integrating Sphere Spectrophotometry method [5] is relatively mature and quite widely deployed. This method measures the reflected light, transmitted light and unscattered light and the result is used to calculate the tissue optical properties. In *ex-vivo* measurement, however, the impact of water loss can be significant so these measurements do not always translate well into an *in-vivo* setting. For *in-vivo* measurement, there are methods like Diffuse Reflectance Spectroscopy (DRS) [6] and Spatial Frequency Domain Imaging (SFDI) [7]. DRS determines the optical properties of opaque samples by measuring the reflected light over a range of wavelengths. The light is transmitted through an optical fiber to illuminate the sample. The reflected light is captured by a probe and passed into a spectrometer. Monte Carlo simulations are often used for data interpretation. SFDI projects multiple patterns with well-defined spatial frequencies onto the sample, a camera then captures the reflected light normal to the sample. The amplitude and phase information of the intensity modulation of the reflected light is extracted from the collected images, and the optical properties of the sample are calculated through light transmission models such as diffusion approximation or Monte Carlo simulation. Although DRS and SFDI are well accepted methods for measuring optical properties, both methods require specialist equipment and careful calibration and the recovered values are strongly dependent on the validity of the optical model and are sensitive to ambient light [8]. Moreover, in most cases these methods do not reveal the anisotropy factor, *g*.

There have been several works aimed at improving the extraction of tissue optical properties using machine learning, often in combination with Monte Carlo simulations. One of the very earliest works reporting the use of neural networks to recover optical properties was presented in 1992 where the errors in the recovered properties were of the order of 45% which is perhaps hardly surprising since as a single hidden layer with 4 neurons was used [9]. Another early paper using a single layer neural network was published in 1994 [10], a principal aim of this paper was to use a network to speed up the very slow simulations available at the time. The network was trained on values from diffusion theory and most of the optical properties were recovered to better than 10% error. The limitation of the training in the diffusion regime meant that thin samples less than the mean free path could not be recovered and the values of anisotropy factor and scattering coefficient were always combined into the reduced scattering coefficient. Generally, more recent papers use Monte Carlo simulation which provides a statistical solution to simulate light propagation based on the tissue properties such as thickness, refractive index, scattering and absorption coefficient, and anisotropy, to provide the ground truth needed for training [11]. The approach is to use Monte Carlo simulations to generate large amounts of data to train neural networks. The trained neural network can then predict the optical properties of biological tissues by analyzing the distribution of the transmission and/or backscattered photons, for example, from the Monte Carlo simulations. As early as 2009 Warncke *et al.* [12] recognized that the limitations of the diffusion approximation could be potentially addressed by using neural networks to improve data recovery from scattering media. These authors used a fiber optic delivery and detection system to generate the results and used the outputs to recover the parameters. Each simulation contained 1 million simulated photons and 3000 data sets were used. A simple network with no more than three hidden layers was used, presumably due to the limitations in machine learning at the time. The approach was used to predict $\mu_a$ with a 28% RMSE and $\mu_s'$ with a 9% RMSE. In 2011, Zhang *et al*., [13] trained a Genetic Algorithm Optimized Back Propagation (GA-BP) neural network and achieved accuracy for the parameters of around 2%. In 2013 Jager *et al*. processed Monte Carlo simulation results, training different networks for different regimes, with an accuracy of 6% for the reduced scattering coefficient and 3% for absorption [14]. In 2021, Hokr and Bixler [11] trained a fully connected neural network using around 300,000 Monte Carlo simulation results for each data set, with 2.5 x$10^5$ photons simulated in each data set. The neural network has one input layer with 21 neurons, three hidden layers with 150 neurons and one output layer with three neurons. The input of the neural network was formatted into 21 input moments. The relative errors were about 15% for $\mu_a$ and 30% for $\mu_s$. It should be pointed out that the range of the input parameters used was exceptionally wide. Chang and Pramanik [15] combined physical insight into machine learning and demonstrated that this gave an improvement in the prediction accuracy achieved.

In the present work we use Monte Carlo simulations to provide the ground truth and try to incorporate data that encompasses some of the essential physics of the propagation process. Moreover, since our input data is in the form of two dimensional maps we utilize a more advanced deep learning network RESNET-50 [16] to recover the data. Compared to [11] we apply the image data at an additional plane as well as a network more attuned to extraction of parameters from image data. With this additional plane we aim to supply the network with information on the exit angle of the photon and not simply exit position.

The network was trained with 37500 data sets randomly sampled over the range of reduced scattering coefficients and absorption coefficients. 7500 samples were used for both validation and testing. The range of values chosen to recover the parameters was taken from values likely to be seen with real tissue samples [4], [17]. The range of $\mu_s'$ was between 0.5 and 2.8 mm$^{-1}$ $\mu_a$ values were between 0.01 and 1.65 mm$^{-1}$ $g$ was between 0.8 and 0.99. The sample thickness used in the simulations was 1.18 mm which meant that for some of the reduced scattering coefficients the thickness was within the transfer mean free path and some were outside. This is discussed in more detail in the Summary and

Conclusions.

A very significant aspect of our work is that each dataset only contained 40000 photons compared to 250000 in [11] and $10^6$ in [15], in addition we used far fewer datasets. We believe that the RESNET architecture is particularly well suited to finding patterns in the data and ignoring noise, so we can extract more useful information for each simulated photon.

In previous works authors used data from a single plane conjugate with the sample surface-possibly after conversion of the data to moments [11], however, when considering the physics of the scattering process it is intuitively likely that the both the angle of the emerging photons as well as the intensity distribution give important information about the scattering properties. To this end we trained the network on different inputs, selected to consider exit angle and position separately and also combined, and assessed their relative performance.

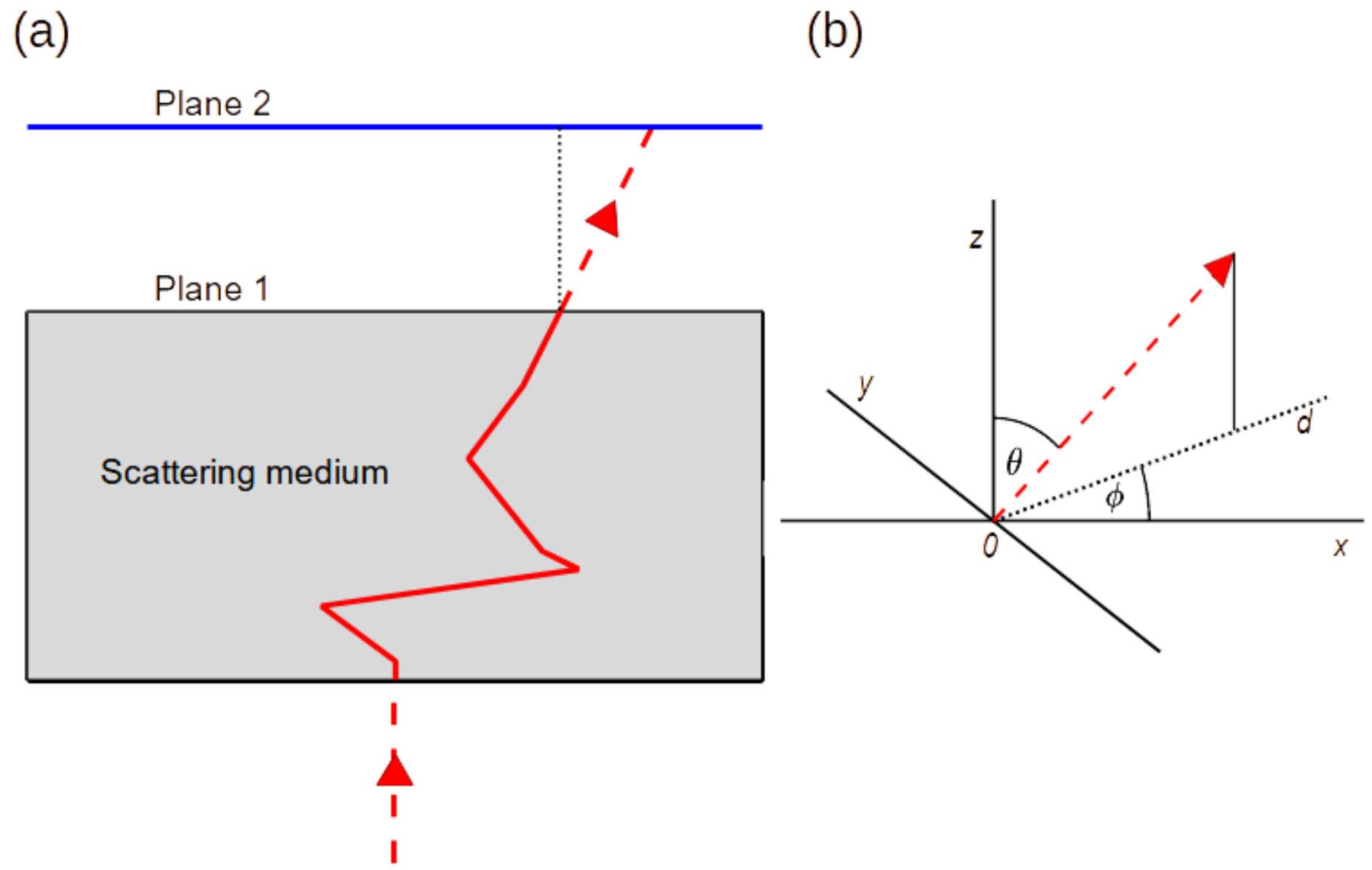

Figure 1 (a) shows a planar view of the random walk of a photon passing through the scattering medium. The path is shown in solid red in the medium and dashed outside where it does not deviate (b) coordinate system defining angles that photon emerges.

Figure 1 shows one particular example of the path of a single photon passing through the scattering medium before emerging. The scattering, of course, occurs in all three dimensions, so the expectation of the intensity distribution emerging from the sample is distributed evenly around the azimuth, $\phi$. The angle of emergence is shown by the dashed red line in Figure 1b and $\phi$ is the angle of the Ozd plane (azimuth) to the *x*-axis and $\theta$ is the angle to the *z*-axis.

To investigate the effectiveness of different metrics to recover the scattering properties we tried different combinations of input parameters. These are given below:

1. Output angles represented as a two dimensional density map plotted as $\phi$ and $\theta$, this is 'Angle' in the tables.

2. Two dimensional Intensity density at plane 1 of Figure 1a. (Surface)
3. Two dimensional map with intensity represented as a single radial density and a combined angle parameter that combines both $\phi$ and $\theta$, namely $\sin\phi\cos\theta$. (Angle+Position in the tables)
4. Two dimensional intensity densities at *two* planes 1 and 2 separated by 5 mm. (2D_5 in the tables)

In practice, measuring the angle is not so straightforward without the use of some wavefront sensor such as a ShackHartmann [18], which generally has limited spatial resolution. For this reason we used two inputs at planes 1 and 2 (Figure 1a) as these are practically much easier to measure and the change in intensity distribution between these two planes would encompass exit angle information. The position of plane 2 was varied and while the results for the scattering coefficients were not very sensitive to the exact separation a value of 5mm between plane 1 and plane 2 gave good results for the reduced scattering parameter (we discuss whether it is optimum for absorption later), hence the notation 2D_5.

In order to make the predictions of the optical parameters we trained three different types of networks (1) to estimate both absorption and scattering ('one model two predictions'), (2) to estimate scattering only and (3) to estimate absorption only. The final two are 'one model one prediction'. Clearly the loss function for each of the cases are slightly different.

For two parameters

1 . $$Loss = \frac{1}{n}\left\{\sum_1^n\left(\frac{\mu_{si}'^{p} - \mu_{si}'^{GT}}{\mu_{si}^{GT}}\right)^2 + \sum_1^n\left(\frac{\mu_{ai}^{p} - \mu_{ai}^{GT}}{\mu_{ai}^{GT}}\right)^2\right\}$$

For scattering only

2. $$Loss = \frac{1}{n}\left\{\sum_1^n\left(\frac{\mu_{si}'^{p} - \mu_{si}'^{GT}}{\mu_{si}^{GT}}\right)^2\right\}$$

For absorption only

3. $$Loss = \frac{1}{n}\left\{\sum_1^n\left(\frac{\mu_{ai}^{p} - \mu_{ai}^{GT}}{\mu_{ai}^{GT}}\right)^2\right\}$$

Where $n$ is the batch size (16 in this case) and the superscripts *p* and GT refer to the predicted and ground truth values respectively. *s* and *a* refer to the reduced scattering and absorption coefficients respectively. The *i* are the individual values in the batch.

| One Model, Two Predictions | | | | |
|---|---|---|---|---|
| Angle | 4.01% | 4.79% | 4.65% | 0.59% |
| Surface | 3.42% | 3.79% | 3.57% | 0.46% |
| Angle + Position | 2.31% | 2.60% | 2.43% | 0.30% |
| 2D_5 | 2.25% | 2.98% | 2.88% | 0.62% |
| One Model, One Prediction | | | | |
| Angle | 3.55% | 4.76% | 4.76% | 1.04% |
| Surface | 3.23% | 3.60% | 3.43% | 0.36% |
| Angle + Position | 2.69% | 3.70% | 3.12% | 1.70% |
| 2D_5 | 2.14% | 2.28% | 2.26% | 0.12% |

Table 1 shows the estimation errors (calculated as the mean modulus of the relative error) for the reduced scattering coefficient for our networks with different inputs.

| | | | | |
|---|---|---|---|---|
| One Model, Two Predictions | | | | |
| Angle | 9.31% | 12.3% | 12.8% | 1.35% |
| Surface | 5.69% | 7.23% | 6.93% | 1.62% |
| Angle + Position | 3.43% | 4.73% | 4.69% | 1.04% |
| 2D_5 | 6.88% | 9.65% | 9.89% | 2.09% |
| One Model, One Prediction | | | | |
| Angle | 11.2% | 12.0% | 11.9% | 0.59% |
| Surface | 5.11% | 6.14% | 6.14% | 0.51% |
| Angle + Position | 3.66% | 4.86% | 4.54% | 0.93% |
| 2D_5 | 8.30% | 9.69% | 9.71% | 1.09% |

Table 2 shows the estimation absorption coefficient (calculated as the mean modulus of the relative error) for our networks with different inputs.

In order to validate the capacity of the network for each of the 4 types of input parameters we trained 8 networks for each set of input parameters and evaluated their performance on the test set. The reason to perform these tests was primarily to ensure that the training was consistent, which appears to be the case since the standard deviation of the errors between networks is much smaller than the mean value. Moreover, we trained networks to recover both the reduced scattering and absorption separately (2 and 3 above) and simultaneously (1). There is no conclusive evidence that the network performs better when it only has a single parameter to retrieve.

It is clear that measuring the angle alone gives the poorest reconstruction of the reduced scattering and absorption, whereas the angle and position gives the best result for both parameters. As explained earlier our rationale for using the two plane surface measurement is that it contains similar information to 'angle and position'; this seems to be borne out for the reduced scattering measurement where it performs comparably to 'angle and position'. For absorption the 'angle and position' outperforms all measurements but the two plane measurement performs worse than a single plane for absorption. This result was a little surprising and is the subject of investigation. We believe that one of the most important parameters for the network is the number of photons emerging and the field of view of the second plane is such that many of the emerging photons are not collected (over a large range of simulated values we can lose as many >5% photons propagating from plane 1 to plane 2). We are examining the effect of reducing the separation between planes 1 and 2 to see if this improves the situation.

**Summary and conclusions**

The present paper presents some preliminary results showing the utility of a general purpose network RESNET50 for the extraction of optical parameters. The network recovers results with comparable or better accuracy relative to previous bespoke networks and recovers the results with a much smaller number of input photons in the training set compared to previous works. In ongoing work we will examine how changing the photon numbers impacts the accuracy of the reconstruction.

In addition, we explored the benefits of using different input parameters in the training set. In general, using both angle and radial exit position gave the best results. We used two planes of intensity distribution as a proxy for both angle and intensity distribution and showed that combining two planes gave considerably better results for reconstruction of the reduced scattering parameter. The use of the second plane, if anything, degraded the accuracy of the estimate of the absorption coefficient. We believe this is because the second plane has not collected all the photons emerging at plane 1, we will report on this effect later publications.

We employed different networks to extract absorption and scattering separately and also both together. It appears that there is no obvious degradation in performance when the model is asked to extract both parameters simultaneously. This is not unexpected as the complexity of the RESNET 50 network is such that we would expect that it could easily configure itself to complete both tasks.

In the present work, we have tried to extract the reduced scattering coefficient and the absorption coefficient, in future work we see how well the scattering coefficient and the anisotropy factor may be extracted separately. We believe this study will give more insight into the process by which the photons are randomized. This is the reason the sample thickness was chosen so that some of the cases were within the transport mean free path (TMFP) and others were well outside this value. The TMFP gives a measure of the randomization of the outgoing photons, within the TMFP there are many unscattered photons emerging, whereas the number diminishes exponentially so that at 10 TMFPs the number of photons is less than 1 in $10^4$. Well beyond TMFP, we do not expect the scattering and the anisotropy to be separable. It is possible that performance can be optimized by two networks, i) tuned to operate in the small scattering regime where $\mu_s'$ may not fully describe the scattering process and ii) operating in the large scattering regime where scattering and anisotropy factor are not separable.

In summary, the RESNET architecture provides a convenient and powerful means to extract optical parameters and compared to previous literature we can extract parameters with a relatively small input training set, and by incorporating angle and position information the accuracy of predicting the scattering parameters is improved.